\documentclass[runningheads]{llncs}
\usepackage[T1]{fontenc}
\usepackage{graphicx}

\usepackage{array}
\begin{document}
\title{Monero’s Decentralized P2P Exchanges:\\ Functionality, Adoption, and Privacy Risks}
\titlerunning{Monero’s Decentralized P2P Exchanges}
%

\author{Yannik Kopyciok \inst{1}\orcidID{0009-0006-4257-4151} \and
Friedhelm Victor\inst{3}\orcidID{0000-0001-8329-3133} \and
Stefan Schmid\inst{1,2}\orcidID{0000-0002-7798-1711}}
\authorrunning{Kopyciok et al.}
%
\institute{TU Berlin, 10587 Berlin, Germany \and Weizenbaum Institute, 10623 Berlin, Germany \and
TRM Labs, San Francisco, USA \\
\email{\{kopyciok,stefan.schmid\}@tu-berlin.de} \\
\email{friedhelm@trmlabs.com}}

%
%

%

\maketitle     

\begin{abstract}
Privacy-focused cryptocurrencies like Monero remain popular, despite increasing regulatory scrutiny that has led to their delisting from major centralized exchanges. The latter also explains the recent popularity of decentralized exchanges (DEXs) with no centralized ownership structures. These platforms typically leverage peer-to-peer (P2P) networks, promising secure and anonymous asset trading. However, questions of liability remain, and the academic literature lacks comprehensive insights into the functionality, trading activity, and privacy claims of these P2P platforms.\\
In this paper, we provide an early systematization of the current landscape of decentralized peer-to-peer exchanges within the Monero ecosystem. We examine several recently developed DEX platforms, analyzing their popularity, functionality, architectural choices, and potential weaknesses. We further identify and report on a privacy vulnerability in the recently popularized Haveno exchange, demonstrating that certain Haveno trades could be detected, allowing transactions to be linked across the Monero and Bitcoin blockchains. We hope that our findings can nourish the discussion in the research community about more secure designs, and provide insights for regulators.
\keywords{Decentralized Exchanges \and Cryptocurrency \and Privacy.}
\end{abstract}

\section{Introduction}
\label{sec:introduction}
Centralized cryptocurrency exchanges have processed nearly three trillion dollars in trading volume in December 2024\footnote{\url{https://www.theblock.co/data/crypto-markets/spot}}, enabling the exchange of fiat and cryptoassets. The vast majority of these exchanges operate within regulatory constraints. For example, in most jurisdictions, financial institutions are required to implement Know Your Customer (KYC) procedures to verify user identities, and monitor transactions for suspicious activity such as money laundering or various types of illicit monetary flows. The Financial Action Task Force's (FATF), an intergovernmental organization developing policies to combat money laundering and terrorist financing, has proposed a ``Travel Rule'', recommending that countries adopt a \emph{de minimis} threshold of 1,000 USD/EUR for virtual asset transfers, above which Virtual Asset Service Providers (VASPs) must collect and share detailed information about the originators and beneficiaries of these transactions~\cite{FATF2021}. In the European Union, the Markets in Crypto-Assets Regulation (MiCA)~\cite{EU2023MiCA}, which fully came into force on December 30, 2024, establishes comprehensive rules for cryptoasset issuance and services, including mandatory authorization and supervision of crypto-asset service providers.

Privacy-focused cryptocurrencies, with Monero (XMR) as the most prominent in 2024, appear to be at odds with prevailing regulations.
Their privacy-enhancing technologies can render transaction monitoring largely ineffective. Although the reason is often not stated explicitly, it is likely that regulations have led to the delisting of Monero from most popular centralized exchanges: OKX Korea and UpBit (2019) followed FATF guidelines, while Bittrex (2021) and Huobi (2022) cited transparency concerns. Binance (2024) removed XMR over compliance conflicts, and Kraken phased out XMR in the European Economic Area at the end of 2024. While there exist very large DEXs on smart-contract supporting blockchains like Ethereum, Binance Smart Chain and Solana, Monero does not supported them.
The combination of regulatory pressure and the disappearance of exchanges has pushed users toward decentralized exchanges (DEXs) and peer-to-peer (P2P) trading platforms supporting Monero. However, popular P2P platforms with centralized operators, such as LocalBitcoins (2023) and LocalMonero (2024), have already shut down.

In response, a new wave of decentralized P2P trading platforms has emerged, each exploring different architectural models to facilitate Monero trading. These platforms differ in network design, trade execution, dispute resolution, and governance but share the common goal of minimizing centralized control.
As these platforms gain traction, questions arise regarding their functionality, and privacy. How exactly do they operate? Do they still rely on centralized components? And what can be observed regarding their privacy guarantees?

\textbf{Contributions:} In this paper, we provide a systematization of the landscape of P2P decentralized exchanges in the Monero ecosystem (cf~Section~\ref{sec:dex_landscape}), covering Haveno, Bisq, BasicSwapDEX, and the COMIT protocol, used e.g. by UnstoppableSwap. For each exchange, we study their functionality, describe the network architecture, the trade protocol, fees, dispute resolution, governance and operational details and potentially remaining centralized components which can also pose privacy risks.
In Section~\ref{sec:haveno}, we examine Haveno in more detail, and unveil a privacy weakness that allows discovering Haveno-related Monero transactions, and in some instances their trade counterpart on transparent blockchains, for example on the Bitcoin blockchain.
Finally, we discuss our findings in Section~\ref{sec:discussion}, focusing on what centralized network components remain, including the aspect of arbitrators and governance structures. We hope that our findings contribute to the ongoing research discussion on more secure designs while also serving as a foundation for addressing regulatory concerns surrounding these P2P exchanges.

\section{Background \& Related Work}
\label{sec:background}
In this section, we outline relevant background knowledge and related works.
Decentralized exchanges provide an alternative to centralized trading platforms by enabling cryptocurrency transactions without intermediaries. While most research on DEXs has focused on automated market maker (AMM)-based exchanges such as Uniswap \cite{angeris2019analysis,Xu2023sokdex,auer2024technology}, P2P DEXs rely on direct interactions between users without the need for smart contracts. In contrast to centralized P2P trading platforms like LocalMonero, which operated under a centralized website, their decentralization extends to the operations, frequently featuring a standalone application, without a centralized website that needs to be accessed. P2P exchange systems differ significantly in their trade execution, governance models, and degree of decentralization. Bisq is one such P2P exchange that facilitates non-custodial trading on the Bitcoin blockchain that has been studied extensively by Hickey and Harrigan~\cite{hickey2022bisq,hickey2020bisq}. They found that its reliance on on-chain transactions makes it susceptible to blockchain analysis techniques such as address clustering, which has been used to trace Bisq trading activity and governance participation through its BSQ token. Haveno, a fork of Bisq designed for Monero, inherits much of Bisq’s architecture but operates within Monero’s privacy framework, aiming to solve some of Bisq's privacy issues.

Critical to decentralized trade execution are atomic swaps, which allow two parties to exchange assets without trusting an intermediary. Hash Time-Locked Contracts (HTLCs) were initially proposed for this purpose and remain a common mechanism in swap implementations, as seen in the COMIT protocol~\cite{hoenisch2021atomic}. However, HTLCs rely on scripting capabilities unavailable in Monero and similar privacy-focused cryptocurrencies. To address this, recent advancements in cryptographic primitives such as adaptor signatures \cite{thyagarajan2022universal} and swap mechanisms that do not rely on locking mechanisms \cite{hoenisch2022lightswap} have extended atomic swap compatibility to non-scriptable blockchains. These  methods may allow for additional privacy-preserving exchange protocols to emerge in the near future.

Monero is designed to obfuscate transaction details using stealth addresses, Ring Confidential Transactions (RingCT), and decoy-based input selection. Unlike transparent blockchains such as Bitcoin or Ethereum, Monero transactions do not publicly reveal sender, receiver, or transferred amounts. However, research has shown that Monero is not entirely immune to analysis. The output merging (or co-spend) heuristic remains a key method for linking Monero transactions, as inputs controlled by the same entity can sometimes be identified when spent together in a single transaction \cite{moser2018empirical}. Cross-chain analyses have also allowed for the deanonmization of input rings~\cite{hinteregger2019short}. Furthermore, while prior research on Monero’s privacy has largely focused on protocol-level weaknesses, more recent efforts have explored application-level vulnerabilities \cite{hammad2024monero}. Cross-chain transactions remain a underexplored risk, as interactions between Monero and transparent blockchains may inadvertently weaken its privacy guarantees~\cite{Goodell2024}.

Cross-chain swap platforms supporting Monero have also been analyzed in terms of their privacy implications. ShapeShift, an early example of an off-chain swap service, was studied by Yousaf et al., who traced transactions revealing user behaviors \cite{yousaf2019tracing}. More recently, similar analyses have been conducted on Evonax, where trading flows have been obtained~\cite{brechlin2025buy,brechlin2024buy}. As decentralized Monero exchanges like Haveno grow in adoption, it becomes increasingly important to investigate whether they exhibit privacy weaknesses.

\section{Monero's P2P DEX Landscape}
\label{sec:dex_landscape}
This section presents a systematization of prominent decentralized P2P exchanges along various dimensions. We will present key insights in the areas of: network architecture highlighting the P2P infrastructure and potential centralized components, trade protocol implementations including the implemented swap mechanisms, fee structure, and dispute resolution frameworks, operational prerequisites to participate on the DEX, and governance structures that regulate protocol modifications and ecosystem development. 
We study Haveno, Bisq, BasicSwapDEX, and COMIT as used by UnstoppableSwap.
After detailing the exchanges, we summarize our findings in Table~\ref{tab:dex-comparison} at the end of this section.

\begin{figure}[htb]
    \centering
    \includegraphics[width=1\linewidth]{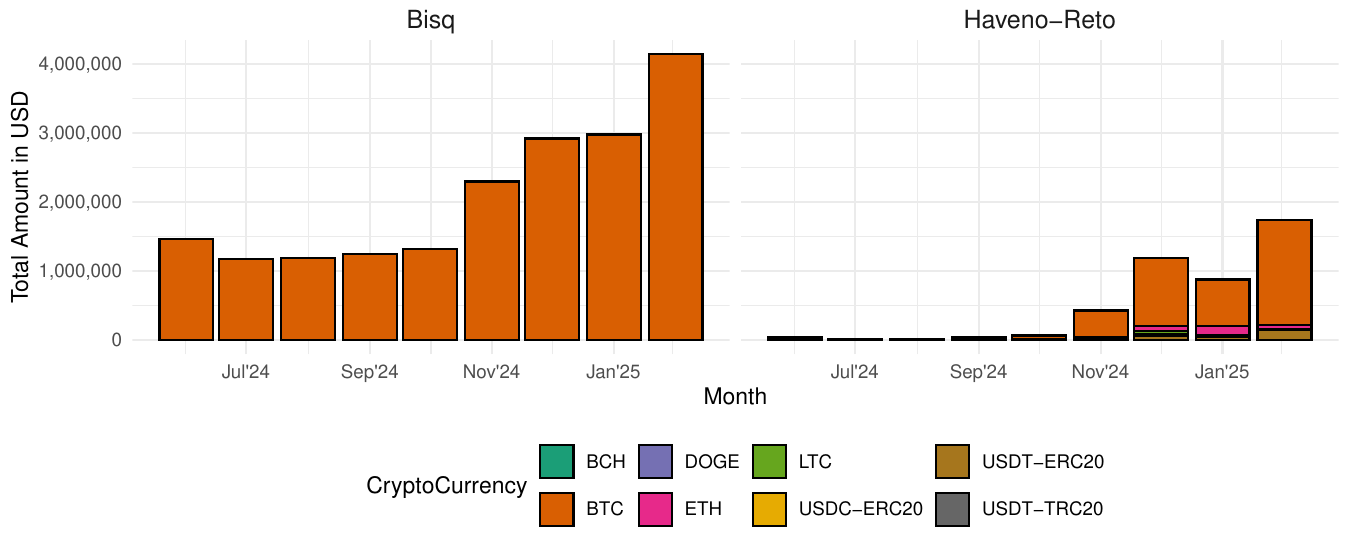}
    \caption{Bisq and Haveno-Reto XMR trading volume against various cryptocurrencies. Bisq only supports XMR/BTC, and has significantly higher volume, but Haveno-Reto (Retoswap), is quickly growing and has approached nearly USD 2 Million in February 2025. While Haveno-Reto supports other XMR Trading Pairs, BTC is the most popular.}
    \label{fig:retoswapVolume}\label{fig:bisqVolume}
\end{figure}

To start, we briefly want to assess popularity of these exchanges. We can only assess historical trade activity for Bisq and Haveno, whereas for BasicSwapDEX and COMIT only a list of offers is obtainable at any given point in time. Figure~\ref{fig:bisqVolume} compares Bisq's with Haveno's Monero trading volume over the past 8 months, when Haveno started seeing usage. On both platforms, BTC is the preferred cryptoasset to trade with Monero. While Bisq is still leading in terms of trading volume, Haveno is quickly rising in popularity, and offers more Monero trading options besides cryptocurrencies, which are not displayed in the chart.

\subsection{Haveno}
Haveno \cite{haveno2025} is an open source project, forked in 2023 from the Bisq \cite{bisq2025} exchange project. Haveno implements a P2P exchange platform based on Tor and Monero and offers a platform to exchange various crypto- and fiat-currencies with Monero as its base currency. No custom or colored token is involved and trades are purely built on Monero's multi-signature wallet implementation. 

\textbf{Network Architecture:\label{havenoNetArch}}
Haveno's core P2P components \cite[docs/deployment-guide.md]{haveno2025} include seed nodes, arbitrator nodes, price nodes, and client nodes. Seed nodes serve as the network backbone, managing state, onboarding new clients, and having their onion addresses hardcoded into binaries. They control seed node management, trade fees, and arbitrator privileges. Arbitrator nodes can register as arbitrators through an authentication mechanism, granting them special privileges. Price nodes run an HTTP service to collect and distribute price data but are not essential for core P2P functionality. Client nodes are regular peers with no special privileges or resource obligations.

Both seed and arbitrator nodes require a local Monero node for security and robustness. All communication occurs via Tor, covering trade interactions from offer creation to execution. Haveno implements a decentralized order book, where offers remain available as long as both the maker and arbitrator are online.

\textbf{Trade Protocol:}
The Haveno trade protocol \cite[docs/trade\_protocol/trade-protocol.md]{haveno2025} makes use of 2/3 multi-signature wallets where the three involved parties are the arbitrator, the entity wanting to buy XMR (taker), and the entity wanting to sell XMR (maker).
During the process of a trade, up to five transactions are created where two of them are penalty transactions in case a party deviates from the protocol. If a taker accepts an offer, the tradable funds and a security deposit from both parties are sent to the multi-signature wallet. After this, the trade must be completed within 24 hours. The taker pays the maker with the respective other coin or fiat currency outside of the Haveno application. The transaction is neither invoked by the Haveno application nor is it monitored. Once the funds are confirmed as received by the maker inside Haveno, the security deposit is refunded to the maker and the taker receives the paid funds plus its own security deposit.
\textbf{Trade fee:} Haveno's code base has the option to enable trade fees but no major fork is applying them at the moment of writing. Therefore, only transactional fees on the respective chains have to be paid.
\textbf{Dispute Resolution:} Haveno offers a chat function that can be used to directly communicate between maker and taker. If an issue is not solvable between the two peers, the arbitrator can be summoned. As the arbitrator is in control of one of the signatures needed for the 2/3 multi-signature wallets, he will assess the dispute and handle accordingly.

\textbf{Governance:}
The Haveno network is dependent on an entity running the initial seed nodes and arbitrators providing arbitrator nodes. Haveno states that they "do not endorse any networks at this time" \cite{haveno2025}. A live mainnet fork of the Haveno repository is currently in active use. It is the earliest and most notable instance named RetoSwap \cite{retoswap}, a.k.a. Haveno-Reto, which has been operational since May 14, 2024.

\textbf{Operational Details:}
Using the Haveno exchange requires only a desktop application and a small amount of Monero for the security deposit. While running a local Monero node is recommended for privacy, it is not required to participate.

\subsection{Bisq}
Bisq \cite{bisq2025}, the precursor to Haveno, shares a similar architecture but uses Bitcoin as its base currency, with Monero as an external payment method. It operates without a central authority, relying on a peer-to-peer network and multi-signature escrow for security. A future update, Bisq2 \cite{bisq22025} plans for new trade protocols including atomic swaps, further enhancing flexibility and interoperability.

\textbf{Network Architecture: \label{bisqNetArch}}
The network architecture shares the same key components as the Haveno network with a key difference in their governance structure.

\textbf{Trade Protocol:}
Bisq's trade protocol \cite[Introduction;Security deposit]{bisq2025} is based on Bitcoin's 2/2 multi-signature wallets where both seller and buyer control their respective private keys. To initiate a trade, both parties contribute security deposits to the multi-signature wallet, with the seller additionally depositing the agreed-upon Bitcoin amount. The buyer has a designated timeframe to transfer the payment via the specified method. Upon the seller confirming receipt, funds are released from the multi-signature wallet to both parties according to the trade terms.
\textbf{Trade fee:} In addition to the transactional fees paid on each blockchain, a small trade fee \cite[Trading fees]{bisq2025} applies which can be either paid in the colored Bitcoin token BSQ or directly in BTC.
\textbf{Dispute Resolution:} Bisq provides an integrated chat for peers to communicate if complications arise \cite[Dispute Resolution in Bisq 1]{bisq2025}. In disputed cases, a mediator can be requested to assess potential resolutions. For severe cases where one party disappears or is uncooperative, Bisq implements a pre-signed time-locked delayed payout transaction that activates after a predetermined period based on the payment method. This transaction directs all funds to designated "donation addresses" managed by the Bisq DAO. The cooperating party can then request arbitration, where an arbitrator investigates the case and reimburses the appropriate amount from their personal funds, later requesting compensation from the Bisq DAO.

\textbf{Governance:\label{bisqGov}}
Bisq is governed by a DAO \cite[Introduction to the DAO]{bisq2025} built on the Bitcoin blockchain. The DAO applies monthly voting cycles to determine revenue distribution, strategic decision-making, and ensuring honesty in high-trust roles. High-trust roles in Bisq involve responsibilities critical to the network, including domain ownership and operating key infrastructure such as seed and arbitrator nodes. These individuals must lock up funds which can be withdrawn by the community in case of misconduct. The high-trust roles do not own any part of Bisq but thereby have a certain incentive to keep the network alive and act in the will of the community. \\

\textbf{Operational Details:}
Participating on the Bisq exchange requires only a desktop application. A user will need an initial small amount of Bitcoin for the security deposit but it is not required to operate a local node. 

\subsection{BasicSwap}
BasicSwap (BSX) is a community-driven and open source DEX project developed within the Particl ecosystem \cite{bsx22024}. Particl consists of multiple privacy-focused decentralized applications, with BasicSwap and a decentralized marketplace as its most prominent products. The ecosystem is built upon its own Particl blockchain implementation, which supports the PART cryptocurrency. Similar to Monero, PART applies privacy-enhancing cryptographic techniques including stealth addresses and ring signatures. A distinctive component of Particl's infrastructure is the SecureMessaging (SMSG) network \cite[Under the Hood]{bsx22024}, an end-to-end encrypted peer-to-peer message mixnet inspired by the BitMessage protocol \cite{warren2012bitmessage}.

\textbf{Network Architecture:}
BasicSwap leverages the SMSG network as its underlying P2P network. SMSG functions as a decentralized storage network where each node maintains a temporary copy of every end-to-end encrypted message broadcast within the last 48 hours. Every BSX client operates a SMSG node serving the broader Particl P2P infrastructure that supports all Particl decentralized applications.
A front-end is hosted locally with every running BSX client. Within this interface, users can interact with posted trade offers. The trade offers are broadcast using the SMSG network, enabling each peer to maintain its own copy of the decentralized order book without relying on centralized matching services. The DEX trading functionality is implemented through atomic swap protocols, with the SMSG network handling the necessary communication required to execute these trustless swaps. 
\textbf{Trade Protocol:}
BasicSwap deploys two atomic swap protocols. If both coins support Bitcoin-style scripting, Hash Time-Locked Contracts (HTLCs) \cite{decred2025atommicswaps} can be used, enforcing swaps through on-chain scripts with shared secret hashes and timelocks. If a coin like Monero lacks these capabilities, swaps rely on adaptor signatures \cite{cryptoeprint:2020/1126}, where private key shares and one-time verifiably encrypted signatures ensure atomicity. Users create and accept offers via BSX's graphical interface, choosing which coins to trade and which swap protocol to use when the involved coins support both. The application then automates the swap process, handling secrets, keys, and transaction ID sharing via the SMSG network. \textbf{Trade Fee:} No additional trade fees apply, only network transaction fees. \textbf{Dispute Resolution:} Both protocols ensure swaps either complete or refund both parties without third-party intervention.

\textbf{Governance:}
The Particl network applies a DAO structure \cite[DAO and Network Treasury]{bsx22024}, details on how the DAO governs BSX is not provided but as trades do not include any fee related to the DAO or another third-party. Financial incentives and involvement in the project appear to be independent from Particl based on community participation rather than direct revenue generation.

\textbf{Operational Details:}
Running blockchain nodes for multiple cryptocurrencies is needed and creates substantial resource requirement. Although Bitcoin synchronization can be accelerated using snapshots and Monero can utilize remote RPC endpoints, the initial setup remains time-consuming. Concurrent node operation demands significant computation, storage, and network resources, potentially deterring casual and test users. These known constraints are scheduled to be addressed as the DEX progresses beyond its current beta phase. 

\subsection{COMIT and UnstoppableSwap}
UnstoppableSwap \cite{uscomit2025} emerged as a successor to the plain COMIT protocol adding a graphical interface to perform cross-chain atomic swaps between Monero and Bitcoin. COMIT is a CLI tool developed in accordance to the whitepaper by Hoenisch and del Pino \cite{hoenisch2021atomic}. 

\textbf{Network Architecture:}
UnstoppableSwap implements a maker-taker model for facilitating atomic swaps. Makers run an Automated Swap Backend (ASB) \cite[docs/pages/becoming\_a\_maker/overview.mdx]{uscomit2025} that provides Monero liquidity to the network. The ASB communicates directly with both the Bitcoin and Monero blockchains through intermediary services: monero-wallet-rpc for Monero and electrs for Bitcoin. While makers typically maintain persistent online presence to service swap requests, takers only connect to the network when initiating trades. Although running dedicated Bitcoin and Monero nodes is recommended for enhanced security and reliability, makers can alternatively configure their ASB to use remote blockchain services.
Two discovery mechanisms \cite[docs/pages/usage/market\_maker\_discovery.mdx]{uscomit2025} are implemented. When starting the UnstoppableSwap application a public registry is visible which is community-maintained and lists available makers. A second discovery mechanism is based on the libp2p rendezvous protocol \cite{libp2p2025discovery}. This protocol enables decentralized peer discovery through rendezvous points, dedicated servers that facilitate connections between network participants. Makers register their services under these rendezvous points, and takers can query them to discover available swap providers. While community volunteers maintain several public rendezvous points, anyone can host their own. Users who already know a specific maker's network address, could directly open peer-to-peer connections.

\textbf{Trade Protocol:}
The trade protocol facilitates atomic swaps as described by Hoenisch and del Pino \cite{hoenisch2021atomic} enabling cross-chain atomic swaps between Bitcoin and Monero. The atomic swaps make use of adaptor signatures where the key mechanism is public keys on both chains sharing the same secret key. \\
In the graphical interface takers are lead through four stages: locking the Bitcoin in a 2/2 multi-signature wallet, the maker locking the Monero, followed by the maker redeeming the Bitcoin and revealing the secret needed for the last step of redeeming the Monero.
\textbf{Trade fees:} Fees are not applied as an additional amount but are included in the exchange rate. When setting up the ASB, a price ticker URL is configured that updates exchange rates based on the current market rate, by default provided by the centralized exchange Kraken. Makers can add a spread to this rate acting as a trade fee. A trade fee going to a third party is not considered.
\textbf{Dispute Resolution:} A four step \cite[docs/pages/usage/refund\_punish.mdx]{uscomit2025} safety mechanism is implemented into the trade protocol: cancel, refund, punish, and cooperative redeem. If the maker is not redeeming the Bitcoin, either party can invoke a Bitcoin cancel transaction. This cancels the swap, and the taker must publish a refund transaction to return the BTC to the specified refund address. If no refund is published, the maker can punish the taker and redeem the Bitcoin. Monero becomes refundable once the swap enters the canceled state. In a punish scenario, the maker may still allow the taker to redeem the Monero, but is not obligated to. 

\textbf{Operational Details:}
As a taker, no additional setup besides installing the graphical interface or even only the CLI is needed. Makers are recommended to run local nodes for security and robustness but setting up local nodes is not enforced.

\subsection{Comparative Analysis}

Table~\ref{tab:dex-comparison} highlights key differences among Monero-focused DEXs. Haveno and Bisq use multi-signature trades with arbitrators, while BasicSwap and COMIT rely on atomic swaps. Haveno and Bisq have centralized seed nodes, whereas BasicSwap and COMIT are fully decentralized. Bisq charges fees, while Haveno remains fee-free for now. COMIT is limited to BTC-to-XMR swaps.

\begin{table}[htbp]
    \centering
    \caption{Comparative Analysis of Monero's Decentralized P2P Exchanges}
    \label{tab:dex-comparison}
    \renewcommand{\arraystretch}{1.2}
    \begin{tabular}{m{2.7cm} m{2.5cm} m{2.3cm} m{2cm} m{2.3cm}}
        \hline
        & \textbf{Haveno} & \textbf{Bisq} & \textbf{BasicSwap} & \textbf{COMIT} \\
        \hline
        \textbf{Governance} & -- & DAO & -- & -- \\
        \textbf{Key Trade\newline Mechanism} & 2/3 Multi-Sig.$^{\dagger}$ & 2/2 Multi-Sig. & Atomic Swap & Atomic Swap \\
        \textbf{Trade\newline Settlement} & P2P + arbitrator & P2P + mediator & P2P & P2P \\
        \textbf{Trade Fee} & Not yet & Yes & No & No \\
        \textbf{Centralized\newline Components} & Seed/arbitrator nodes, (Sec.~\ref{havenoNetArch}) & Seed/arbitrator nodes (Sec.~\ref{bisqNetArch}) & No & No \\
        \textbf{Network \newline Communication} & Tor & Tor & SMSG & libp2p, optionally via Tor \\
        \textbf{Supported \newline Assets} & XMR $\leftrightarrow$ * & BTC $\leftrightarrow$ * & ** & BTC $\rightarrow$ XMR \\
        \hline
    \end{tabular}
    
    \smallskip
    \small
    * Universal compatibility, including XMR.\\
    *** BTC, XMR, DASH, LTC, FIRO, PIVX, DCR, WOW, PART.\\
    $^{\dagger}$ Arbitrator controls the third signature.
\end{table}


\section{Haveno Vulnerability and Cross-chain Observability}
\label{sec:haveno}
During our analysis we recognized different potential weaknesses a DEX might be prone to. If the network architecture includes centralized components, it offers a central point of failure potentially bringing the whole network to a collapse.
Centralized components raise legal concerns if critical infrastructure is controlled by a single entity, potentially qualifying as a financial service. Trade protocols also introduce risks beyond security vulnerabilities; their design may create detectable on-chain patterns, undermining privacy guarantees through blockchain analysis.
On-chain observability greatly depends on the underlying cryptocurrency. While Bitcoin provides full transparency regarding wallet addresses and transaction amounts, Monero inherently obscures such information, though timing, transaction structure, and fees are still observable. Each structural characteristic gains in importance when observed in a specific arrangement including temporal observations. For trades involving fiat currencies, while the fiat transactions remain hidden for public observers, they maintain full visibility to the participating financial institutions.

The P2P network is usually used to communicate details before, during, and after trades, mostly only between the involved parties. Nonetheless, messages broadcast to the network can provide orientation for the activity of the application. Furthermore, it adds a time metric if offers, takes, and successful or unsuccessful trades are broadcast to the network.

\subsection{Haveno Data Observability}
Haveno offers multiple distinct features that can be publicly observed within the application, the P2P network, and the Monero blockchain. We will first derive an observable on-chain pattern from the trade protocol and continue to combine the blockchain information with application data. As the Monero blockchain hides most transaction details, trades involving another currency operating on a public ledger like Bitcoin will play a key part to identify both trade partners.

\begin{figure}
    \centering
    \includegraphics[width=1\linewidth]{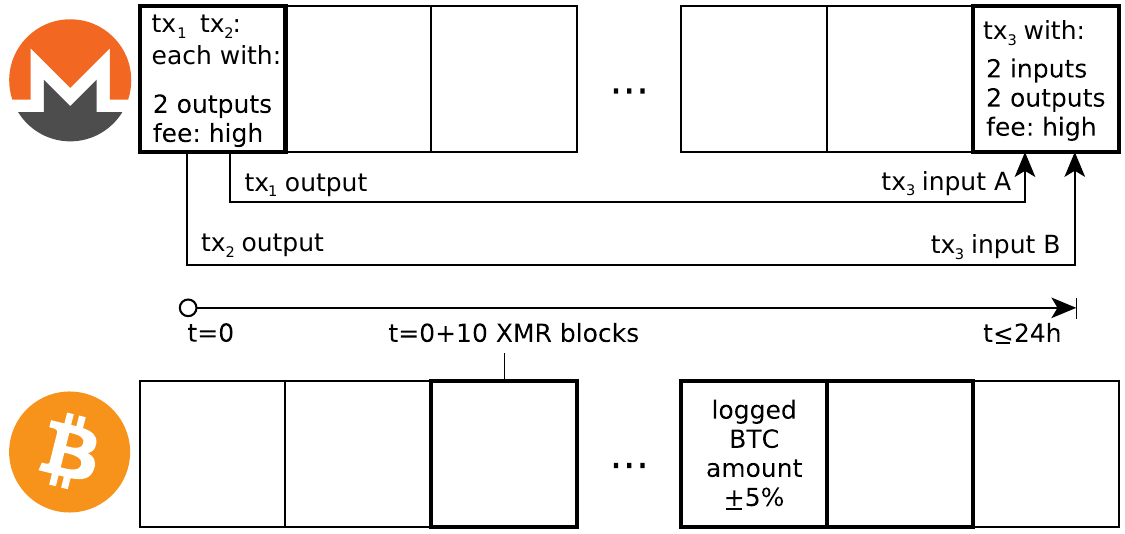}
    \caption{On-Chain Pattern for a standard Haveno trade without a dispute. The two outputs created in the first two transaction are referenced in the final transaction within 24 hours. Within this time window, a counterpart needs to exist on the other trading currency, in this case Bitcoin. Haveno by default uses the same high fee for both transactions, obfuscates the reported trade amount, but broadcasts the completion of the trade immediately via the P2P network.}
    \label{fig:havenoChainPattern}
\end{figure}

\textbf{Trade Protocol Analysis:\label{havenoTradeProtocolAnalysis}}
According to the trade protocol, a trade includes three transactions. Two lock transactions as security deposits and one spend transaction as payout. Both security deposits are invoked once a trade is taken and broadcast to the network by the arbitrator. The trade protocol allows the taker 24 hours to send the external funds to the seller and once the maker confirms this, the trade concludes with the payout. On-chain, this means that within 24 hours after the lock transactions, one output of each lock will be a ring member for the spend transaction as visualized in Figure~\ref{fig:havenoChainPattern}. Both lock transactions will have two outputs, and the spend transaction will have two outputs and two inputs. It is possible that the two security deposit transactions are split and written to two blocks if the the network delays a transaction or a block reached its limit. But as the arbitrator verifies the trade details and transactions before broadcasting everything, most lock transactions are written to a single block. Finally, Haveno transactions use an increased fee structure. Although the fee is one of the standard fees commonly used by third party Monero applications, the high nature still rules out a majority of transactions that use the lower standard fee. See Figure~\ref{fig:havenoChainPattern} for an illustration of the on-chain pattern.

\textbf{Trade Statistics:}
Haveno offers historical trade statistics. The amounts involved in a trade are obfuscated by +-5\% setting the frame for potential transactions on the Bitcoin blockchain too large to draw relevant conclusions. However, during the obfuscation the exchange rate stays accurate. Therefore, by reversing the heuristic, we can extract every Bitcoin transaction that fits into the range of the obfuscation time window. As the Haveno interface allows the XMR to include a decimal number with a maximum of four digits after the point the number of potential Bitcoin transactions is already greatly reduced. 

\textbf{TradeLogger:}
A second obfuscation in the trade statistics happens on the trade timestamp. The timestamp is shifted by some time between zero and 24 hours. However, the Haveno application immediately broadcasts the trade to the statistics once the trade is successfully completed. There might be delays in the network which unintentionally obfuscate the timestamp but typically, this notification in the network gives us the time when a trade has been completed. Combined with our trade protocol, this gives us roughly the timestamp of the XMR spend transaction, the payout. Therefore, we can narrow down external transactions from a 24 hours window to the time frame we get of the observed on-chain pattern.
\subsection{Results}
To test our findings, we logged Haveno trades for two weeks and executed five test trades within the observation period. For all five transactions, we successfully identified all XMR transactions. Furthermore, our analysis correctly isolated a subset of BTC transactions including the real BTC payment transaction.

\newpage
\noindent\textbf{Datasets:}
For our evaluation we collected three datasets: 
\begin{itemize}
    \item Monero mainnet blockchain transactions from 2025-01-21 until 2025-02-03
    \item Bitcoin mainnet blockchain transactions from 2025-01-21 until 2025-02-03
    \item RetoSwap Trade Statistics
    \item RetoSwap trades logged from 2025-01-22 until 2025-02-03
\end{itemize}
The data includes 344 trades logged during this period, 66 of which used Bitcoin as the exchange currency. Our analysis will focus on these Bitcoin trades.

\textbf{Monero Blockchain Analytics:}
To match our logged trades we started with scanning the Monero blockchain for potential swaps. As described, a swap includes two lock transactions and a single spend transaction.
As the spend transaction provides the link to the lock transactions, we searched the chain in reverse starting with every possible spend transaction. For each possible spend transaction, we collect the transaction ID and all ring members of both inputs. The lock transactions must have been made within 24 hours before this block and to take into account the possibility that the lock transactions are split, we allow the lock transactions to be in neighboring blocks.

Within the given time frame of 14 days, we find 671 potential swaps that fit with the observable Monero pattern. For comparison, within these 14 days (block 3,330,482 to 3,339,845), 371,206 transactions are written on the Monero blockchain with 14,666 matching the structural characteristics of a spend transaction. Only for 671 transaction, we can match potential corresponding locks and include them as potential swaps. 

\textbf{Haveno Trade Matching:}
We continue by combining the found swaps with the trades logged from the RetoSwap application. The spend transaction from the potential swap should be in a block that was mined around the time the trade was broadcast. We therefore match potential swaps when the spend was mined within ten minutes after or one minute before the trade was logged to account for network delays. We thereby found for 65 of the logged Bitcoin trades 98 potential swaps. Multiple swaps for a single trade are expected and not ultimately decide the matching accuracy as it mainly influences the time frame we will consider Bitcoin blocks for each swap. 

\textbf{Bitcoin Blockchain Analytics:}
As the amount of BTC and XMR is obfuscated, we first include a BTC transaction when it is within the possible range. Afterwards, we use the exchange rate to calculate the amount of XMR. If the resulting XMR amount is a real number with equal or less than four digits after the decimal point, we include the Bitcoin transaction as a possible transaction.

\textbf{Matches:}
Taking all possible Bitcoin transactions into account, we match one Monero swap pattern to on average 933 Bitcoin transactions. But the majority of Bitcoin transactions relate to an XMR amount with four digits after the decimal point.
We can reduce the set of possible transactions to a realistic subset by assuming traders usually trade even amounts. Matching amounts divisible by 1, 0.5, 0.25, or 0.1 reduces the candidate set to a size suitable for manual inspection. Using this subset, we find transactions for 50 trades, matching each to an average of 2.5 Bitcoin transactions, with a median of one transaction per swap.

\section{Discussion}
\label{sec:discussion}
Our analysis of decentralized exchanges revealed several key differences. We observed varying network and governance structures across platforms, alongside considerable diversity in available trade protocols. These protocols facilitate varying cryptocurrency compatibility and potentially introduce vulnerabilities in cross-chain transaction privacy.

\textbf{Central Network Components:}
Haveno's and Bisq's network architecture requires seed node infrastructure to be deployed before arbitrator nodes can register, and arbitrators have to be registered and online for peers to start trading. These components create potential centralization vectors. The disruption of seed nodes could render the network inoperative, while compromised arbitrator nodes would block trading functionality. Multiple seed and arbitrator nodes can and should exist to keep the network live 24/7 and simultaneously create a form of decentralization within the seed and arbitrator node infrastructure. Nonetheless, not every node and therefore not every peer is equal within the network, therefore challenging the claim that there is no central authority. If the seed nodes are taken off the grid, the network cannot exist.
For BasicSwap, the integration with Particl provides robustness by eliminating centralized points of failure but it also creates some dependence and could present scalability challenges as the ecosystem grows. Decentralized applications within the ecosystem share a security model where threats targeted at either the Particl blockchain or the SMSG network would potentially impact all applications. However, this shared security model also means improvements and a growing network will benefit all applications.
The P2P network architecture implemented in the COMIT protocol mitigates single points of failure through its distributed and lightweight design. Trading and communication is peer to peer. Only to discover makers centralized services are beneficial. The network has two distinct node types but each peer remains equal. Everybody can deploy an Automated Backend Server and act as a maker, and everyone can take the role of a taker.

\textbf{Arbitrator Involvement:}
The integration of arbitrators and third-party mediators in certain DEX implementations presents significant implications for transaction security and privacy.
Whereas BasicSwap and COMIT implement direct atomic swaps between two participants, Haveno and Bisq incorporate arbitration mechanisms that necessitate third-party involvement. The arbitrator has to be online for a peer to register the offer, to execute the trade, and for dispute resolution. A major difference between Haveno and Bisq is the involvement in the multi-signature wallet. Haveno's arbitrators have a third signature, giving them not sole control but still the power to align with either party and release the funds from the wallet. Bisq implements a more decentralized and objective approach where the arbitrator acts more as a mediator, objectively supporting the dispute resolution without actually giving them any power.
Bisq's arbitrators seem to only provide infrastructure governed by the DAO, Haveno's trade protocol could be interpreted as giving the arbitrator actual access to the funds. Nonetheless, both scenarios might raise the question if some kind of financial service is provided by a single entity. This raises the regulatory question of whether arbitrators are facilitators of cryptocurrency exchanges.
Furthermore, as the arbitrator is in possession of the third signature, the arbitrator necessarily requires access to Monero transaction details, challenging the claim of true anonymous peer to peer interaction. The arbitrator will only in dispute cases see the involved Bitcoin addresses but as he has access to the full transaction details including an accurate Bitcoin amount, a blockchain analysis could easily reveal the transaction in question. This raises serious privacy concerns if an arbitrator is not acting in good will and would require putting trust in a third-party.  

\textbf{Governance:}
Governance structures across the examined platforms reveal contrasting approaches to decentralized decision-making, with varying degrees of transparency and stakeholder participation that directly impact platform resilience and ultimately trust.
Bisq and BasicSwap apply a DAO structure outlining treasury, infrastructure, and voting mechanisms. Whereas for Bisq the DAO has to decide on directly relevant issues regarding the DEX like fee structure and node management, BSX DAO is more aligned with the general Particl ecosystem and development as no specific infrastructure for BSX is needed and no trade fees apply.
Haveno states that a live network is neither maintained nor endorsed. RetoSwap is supposed to be an independent structure running its own live network. Nonetheless, it appears that besides running the live network and hosting another website, the project and its repository appears identically.

\textbf{Haveno:}
Haveno has been discussed in greater detail as it evolved to one of the most prominent exchanges in the context of Monero. While strong promises claim privacy with every transaction and independence from any central authority, the current implementation raises uncertainty. Our analysis showed detectable on-chain patterns and weaknesses in the platform that can be exploited to match transactions across chains. These vulnerabilities may be addressed through standardized fee structures that enhance transaction anonymity within larger transaction volumes. While trade statistics provide valuable metrics for users, their network propagation should be obfuscated through mechanisms such as aggregated daily updates to preserve trade privacy.
A more complicated matter is within the missing governance structure. In contrast to the COMIT network, a Haveno network is not functional without an initial setup. Two peers cannot boot respective nodes and trade. Therefore, a decentralized transparent process regarding the decentralized critical infrastructure is necessary to make it really independent from central control. Bisq is an interesting example as it deploys the same architecture with a more transparent governing structure. Nonetheless, even with an existing clear governance, the described involvement of a third party is concerning especially with the motive of using Monero as a privacy first alternative.  

\section{Conclusion}
\label{sec:conclusion}
In this paper, we have provided a systematization of the landscape of P2P decentralized exchanges in the Monero ecosystem, covering Haveno, Bisq, BasicSwapDEX, and the COMIT protocol, used e.g. by UnstoppableSwap.
Our analysis reveals significant gaps between the decentralization and privacy promises of Monero-focused DEXs and their actual implementations. While these platforms offer alternatives to increasingly regulated centralized exchanges, they exhibit varying degrees of decentralization and privacy.  

Haveno and Bisq rely on centralized components such as seed and arbitrator nodes, which introduce resilience and regulatory concerns. Additionally, we demonstrate that Haveno trades leave detectable on-chain footprints, allowing some degree of cross-chain transaction linking. Truly decentralized alternatives like COMIT eliminate these risks but sacrifice user experience and liquidity.  

Governance structures range from well-defined decentralized autonomous organizations to loosely structured community-driven projects, raising questions about long-term sustainability and legal implications. Future work could explore mitigation strategies for privacy weaknesses in DEXs and evaluate emerging cross-chain trading techniques that aim to preserve Monero's anonymity. Our findings contribute to the discussion on secure DEX designs and provide a foundation for assessing regulatory concerns surrounding P2P cryptocurrency exchanges.

\section*{Acknowledgement}
Partially funded by German Research Foundation (DFG), project ReNO (SPP 2378), 2023-2027.

%
%
%
%
%
%
\bibliographystyle{splncs04}
\bibliography{main}
\end{document}